# Structural and transport properties of 4*f* electron doped $Y_{1-x}(Dy)_xPdBi$ topological semi-metallic thin films

Vishal Bhardwaj[a], Niladri Banerjee[b+], Ashok K. Ganguli[c] and Ratnamala Chatterjee[a*]

[a]Department of Physics, Indian Institute of Technology Delhi, Hauz Khas, New Delhi-110016, India

[b]Department of Physics, Loughborough University, Loughborough, LE11 3TU, United Kingdom

[c]Department of Chemistry, Indian Institute of Technology Delhi, Hauz Khas, New Delhi-110016, India

*ratnamalac@gmail.com, [+] N.Banerjee@lboro.ac.uk

We report the effect of 4*f* electron doping on structural, electrical and magneto-transport properties of Dy doped half Heusler $Y_{1-x}(Dy)_xPdBi$ ($x$ =0, 0.2, 0.5, 1) thin films grown by pulsed laser deposition. The Dy doping leads to lattice contraction which increases from 0% for the parent $x$ =0 sample to ~1.3% for $x$=1 sample with increase in Dy doping. The electrical transport measurements show a typical semi-metallic behaviour in the temperature range 3K ≤ T ≤ 300K and a sharp drop in resistivity at low temperatures (< 3K) for all the samples. Magnetotransport measurements and Shubnikov de-Hass oscillations at high magnetic fields demonstrate that for these topologically non-trivial samples, Dy doping induced lattice contraction plays an active role in modifying the Fermi surface, carrier concentration and the effective electron mass. There is an uniform suppression of the onset of superconductivity with increased Dy doping which is possibly related to the increasing local exchange field arising from the 4*f* electrons in Dy. Our results indicate that we can tune various band structure parameters of YPdBi by *f* electron doping and strained thin films of $Y_{1-x}(Dy)_xPdBi$ show surface dominated relativistic carrier transport at low temperatures.

Several members of rare-earth (R)-based RPdBi half Heusler (HH) family are known to exhibit localized magnetism of $R^{3+}$ ions and order antiferromagnetically at low temperature[1,2]. The MgAgAs type crystal structure of these alloys has high cubic symmetry with the absence of inversion symmetry. The cubic symmetry results in *s*-like $\Gamma_6$ and *p*-like $\Gamma_8$ degenerate bands at the $\Gamma$ point around the Fermi level with low density of states, characterizing them as semimetals[3]. The presence of strong spin orbit coupling produced by heavy metallic elements like Bi can invert the $\Gamma_6$ and $\Gamma_8$ bands in RPdBi alloys thus, resulting in a topologically non-trivial electronic structure of these semimetals[4–6]. The RPdBi HH can be classified into either topologically trivial or non-trivial state depending on the lattice density and spin orbit coupling strength of the R element; which further varies with lattice contraction (LC) and effective nuclear charge (Z) as $Z^4$, respectively[2,4–6]. Interestingly, some alloys from this class also show superconductivity at low temperature with critical temperature $T_C < 2K$[2,7–10]. Thus, coexistence of magnetic ordering and superconductivity, along with the topological non-triviality make these semimetals promising candidates to host novel collective excitations like Dirac[11], Weyl[12] and Majorana[13] fermions. These relativistic carriers in the topological semi-metals show high mobility, very high magneto-resistance and dissipationless transport - key features for future spintronic [14,15] and quantum computing applications[16,17].

The Lanthanide series (La to Lu) of rare-earth elements consist of localized 4*f* electrons and most of the elements are found in +3 oxidation state. The magnetic ordering in RPdBi alloys results from the exchange coupling between these localized 4*f* moments. Although yttrium (5d) is a transition element, it is considered as rare-earth element due to its similar chemical properties to lanthanides. Bulk unstrained YPdBi HH system is predicted to exist in topologically trivial state without any band inversion; and due to the absence of *f* electrons, it shows diamagnetic behavior[1,2,4,6,9]. The topologically trivial nature of YPdBi single crystals has been confirmed in recent years using nuclear magnetic resonance [18–21] and electron spin resonance [22,23] techniques. Using DFT, Chadov *et al.* predicted that band structure of YPdBi can be tuned between topologically trivial and non–trivial state by varying the lattice constant and/or spin orbit coupling strength of the material system[6]. Although the lattice could be strained by applying uniaxial pressure, an easier way is to grow epitaxial oriented thin films on carefully selected substrates to induce a lattice strain[9,24–26]. We have shown in an earlier work using experimental and band structure computations that ~3.12% strained YPdBi (~30nm) thin films are topologically non-trivial [9]. In other work, we have also shown the topological non-trivial nature of strained DyPdBi[25,26] thin films. Here we report a systematic study of the electrical and magneto-transport properties of 4$f$ electrons (Dy) doped YPdBi thin films with the aim of characterizing their topological behavior as a function of Dy doping. .

In this work, the (110) oriented Y$_{1-x}$(Dy)$_x$PdBi ($x$ =0, 0.2, 0.5, 1) thin films (~30nm) were grown on single crystal (100) MgO substrates using pulsed laser deposition (PLD) from targets prepared using *rf* induction melting technique. The thin films show strained lattice structure with respect to their bulk samples and the percentage strain decreases from ~3.1% for $x$ =0 sample to ~1.8% for $x$ =1 sample. The lattice constant of thin films uniformly contracts with increase in Dy concentration from~ 6.84Å for $x$ =0 sample to ~6.75Å for $x$ =1 sample which implies a LC of ~1.3% for the $x$=1 sample with respect to the $x$=0 parent sample. The resistivity measurements show the onset of superconductivity-like sharp transitions at low temperatures with extrapolated $T_C$ decreasing with increasing $x$. The magneto-resistance also decreases with increase in $x$ and indicate weak anti-localization effect at low magnetic fields which is fitted to Hikami-Larkin-Nagaoka theory. Shubinkov de-Hass oscillations are observed in high field (>4T) magneto-resistance data and systematic variation of various electronic structure parameters with increase in $x$ is extracted by fitting these oscillations to standard Lifshitz-Kosevich theory. Our experimental results indicate that the *f* electron doping at rare-earth site offers an effective way to modify the electronic structure of the parent YPdBi system.

The polycrystalline bulk samples and thin films of Y$_{1-x}$(Dy)$_x$PdBi were prepared using custom designed *rf* induction melting and PLD techniques as described in ref. [9,26]. The Y$_{1-x}$(Dy)$_x$PdBi thin films of thicknesses ~30nm were grown at optimized in situ substrate temperature of 270ºC on (100) MgO single crystal substrates of dimension ~ 3mm × 10mm with ~ 5nm Ta seed layer. In this work YPdBi, Y$_{0.8}$Dy$_{0.2}$PdBi, Y$_{0.5}$Dy$_{0.5}$PdBi and DyPdBi samples are studied, which are labelled as D0, D2, D5 and DPB, respectively. The X-ray diffraction (XRD) patterns of bulk powder and thin film samples were obtained at room temperature using Panalytical X'pert Highscore x-ray diffractometer using Cu K$_\alpha$ (1.54Å) radiation source. The X-Ray reflectivity (XRR) and rocking curve were measured using same diffractometer. Longitudinal resistivity and magneto-resistance data of thin film sample were measured using the four-probe method with a dc probe current of 50µA in a commercial PPMS system from Quantum Design which can sweep the magnetic field between ±9T at lowest temperature up to 1.9 K.

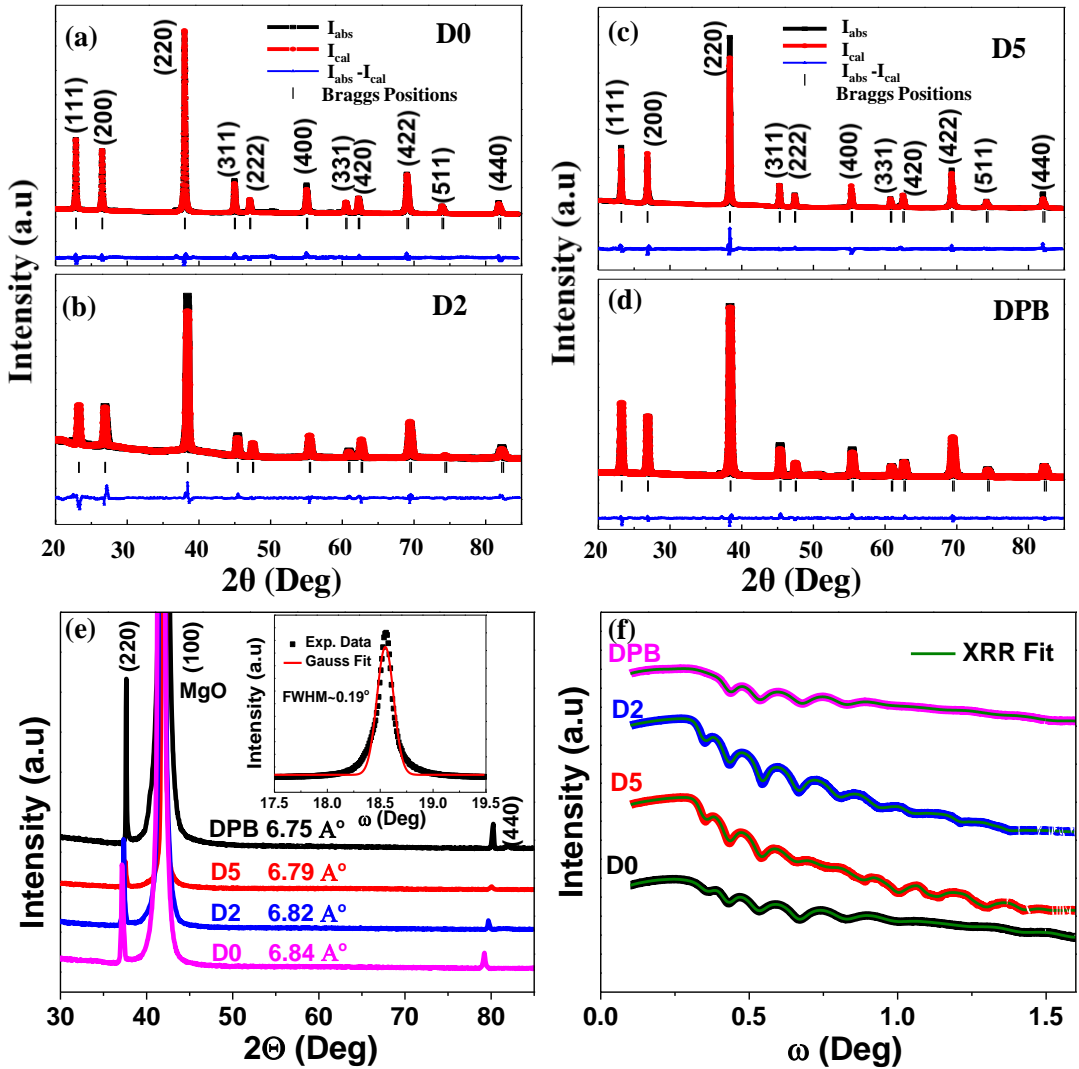

**FIG. 1.** Powder XRD patterns of bulk samples and their Rietveld refinement for (a) Sample D0, (b) sample D2, (c) sample D5 and (d) sample DPB. (e) Gonio-mode XRD patterns of D0, D2 and D5 thin films with [110] oriented planes. Inset shows rocking curve scan of (220) peak of D0 sample. (f) Specular XRR spectra of D0, D2 and D5 samples.

Figures 1(a-d) show representative powder XRD patterns and Rietveld refinements of the bulk D0, D2, D5 and DPB[26] samples. The structure refinement was performed using FullProf software. The theoretically calculated and experimentally observed diffraction patterns are in good agreement with each other as seen from the linear difference patterns represented by blue lines in Figs. 1(a-d). The Rietveld refinements confirm the MgAgAs-type ($C1_b$) cubic crystal structure with $F\bar{4}3m$ space group. The reflections are well-indexed to $C1_b$ structure and lattice constant ($a$) decreases from 6.638Å to 6.631Å with increase in $x$ from 0 (D0) to 1 (DPB), see Table 1 and are consistent with the literature data[27,28]. Figure 1(e) shows the XRD spectra of D0, D2 and D5 thin film samples recorded in Gonio mode (2θ/θ), which indicate the observation of only symmetric reflections[29]. The presence of (220) and (440) Bragg peaks indicate the (110) oriented growth of all samples. The degree of orientation of films is verified by performing the rocking curve (ω-2θ) scan around (220) peak. Inset of Fig. 1(e) shows an example of the ω-2θ scan for the D0 sample. A very small FWHM value ~ $0.19^o$, $0.16^o$, $0.14^o$ and $0.10^o$ is obtained from Gauss curve fit to the rocking curve for samples D0, D2, D5 and DPB respectively[9,29]. The lattice constant ($a_{TF}$) of thin films is calculated as 6.84Å, 6.82Å, 6.79Å and 6.75Å with tensile strain (with respect to bulk samples) ~3.1%, 2.8%, 2.4% and 1.8% for samples D0, D2, D5 and DPB, respectively. The LC% increases from 0% for D0 sample to ~1.3% for DPB sample, see Table 1. The decreasing trend in FWHM of (220) peak from rocking curve also indicate the relaxation of the strain with increase in $x$ from 0 to 1 [30,31].

The thin film growth rate is optimized using XRR measurements. Figure 1(f) shows the experimental and simulated (green lines) XRR curves obtained by segmented fit and generic algorithms. The individual layer thickness ($t_{Ta}$ and $t_{YDPB}$), interface roughness ($\sigma$) and the density of layers ($\rho_{Ta}$ and $\rho_{YDPB}$) are estimated from XRR fitting as shown in Table S1 in supplementary information. We note that all films have low roughness around 1~2nm without any particular correlation with Dy doping. However, the density− $\rho_{YDPB}$ of $Y_{1-x}(Dy)_xPdBi$ layer increases from ~ 8.45 g/cc for D0 sample to ~ 9.71 g/cc for DPB sample which also indicate the increased lattice density resulting from LC with increase in Dy doping.

**Table 1.** Summary of the structural parameters estimated using Rietveld refinement for bulk samples and thin film parameters from Gonio mode XRD data. Atomic positions Y/Dy: x=y=z=0; Bi: x=y=z=1/2; Pd: x=y=z=3/4, α=β=γ=$90^o$, Wyckoff positions: Bi=4a, Y/Dy=4b, Pd=4c and space group 216 (F$\bar{4}$3m).

|  | **D0** | **D2** | **D5** | **DPB** |
|---|---|---|---|---|
| $\chi^2$ | 1.50 | 3.84 | 2.37 | 1.79 |
| **Occupancy** | Y=1<br>Pd=1<br>Bi=1 | Y=0.80<br>Dy=0.20<br>Pd=1<br>Bi=1 | Y=0.50<br>Dy=0.50<br>Pd=1<br>Bi=1 | Dy=1<br>Pd=1<br>Bi=1 |
| (a=b=c)(Å) | 6.638 | 6.635 | 6.633 | 6.631 |
| $a_{TF}$ (Å) | 6.84 | 6.82 | 6.79 | 6.75 |

| FWHM (Deg) | 0.19 | 0.16 | 0.14 | 0.10 |
| Strain (%) | 3.1 | 2.8 | 2.4 | 1.8 |
| LC(%) | 0 | 0.3 | 0.7 | 1.3 |

The electrical and magneto-transport measurements were performed on rectangular samples with dimensions 3mm × 10mm with Ohmic contacts made by using silver paste cured at room-temperature [9,24,26]. The temperature dependent longitudinal resistivity ($\rho_{xx}$) curves show a typical semi-metallic behavior for all samples in temperature range 1.9 K ≤T≤ 300 K and a sharp drop in resistivity is observed at low temperatures for samples D0[9], D2 and D5. It should be noted that although $\rho_{xx}$ drop is sharp, it does not reach zero as these measurements were limited by the base temperature of the cryostat at 1.9 K. We extrapolated the curves to zero resistance and noted the temperature corresponding to 50% of the resistance drop as shown in the inset of Fig. 2(a). If these sharp drops in resistance occur due to the onset of superconductivity, these temperatures would correspond to the superconducting transition temperature. The estimated $T_C$ for D0, D2 and D5 samples are $T_{D0}$ ~ 1.75 K, $T_{D2}$ ~ 1.58K and $T_{D5}$~ 0.94 K, respectively indicating a progressive suppression of this superconductivity like phenomenon with increasing Dy doping. In presence of an external magnetic field, this sharp drop in $\rho_{xx}$ is gradually suppressed in all the samples until it turns fully normal as shown in Fig. S1 and ref.[9] .

Recent studies of superconductivity in single crystal samples of YPdBi and DyPdBi semi-metals establish them as noncentrosymmteric superconductors[2,10]. The $T_C$ for YPdBi was reported ~1.5K by Nakajima et. al.[2] and ~1 K by Radmanesh et. al.[10], whereas for DyPdBi $T_C$ ~ 0.9K has been reported by Nakajima et. al.[2]. In our recent work, we also observed a superconducting state with $T_C$ ~1.25K in (110) oriented strained YPdBi thin films[9]. Therefore the sharp drop in $\rho_{xx}$ of $Y_{1-x}(Dy)_xPdBi$ thin films and disappearance of this transition with increasing external magnetic field most likely indicates the onset of a superconducting state.

The magnetoresistance (MR) of samples is measured in four probe geometry and the direction of field is perpendicular with respect to the plane of the film. We measured MR as a function of temperature down to 1.9 K and the normalized MR% of samples D0, D2, D5 and DPB at 2 K is shown in Fig. 2(b) for comparison. At 2 K the MR% decreases progressively with Dy doping with D0 (black open symbols) showing a MR~32% (at H=±9T) which decrease to ~0.6% (at H=±7T) for the pure DPB sample (blue closed symbols) with the intermediate doped samples D2 (black closed symbols) and D5 (blue open symbols) showing a MR of 32% (at H=±9T) and 5.2% (at H=±9T) respectively (see Fig. 2 (b)). The MR over a wider range 1.9K ≤ T≤ 50K for samples D2 and D5 is shown in Figure S2. All samples show decrease in MR with increase in temperature and the MR disappears at 50K[9].

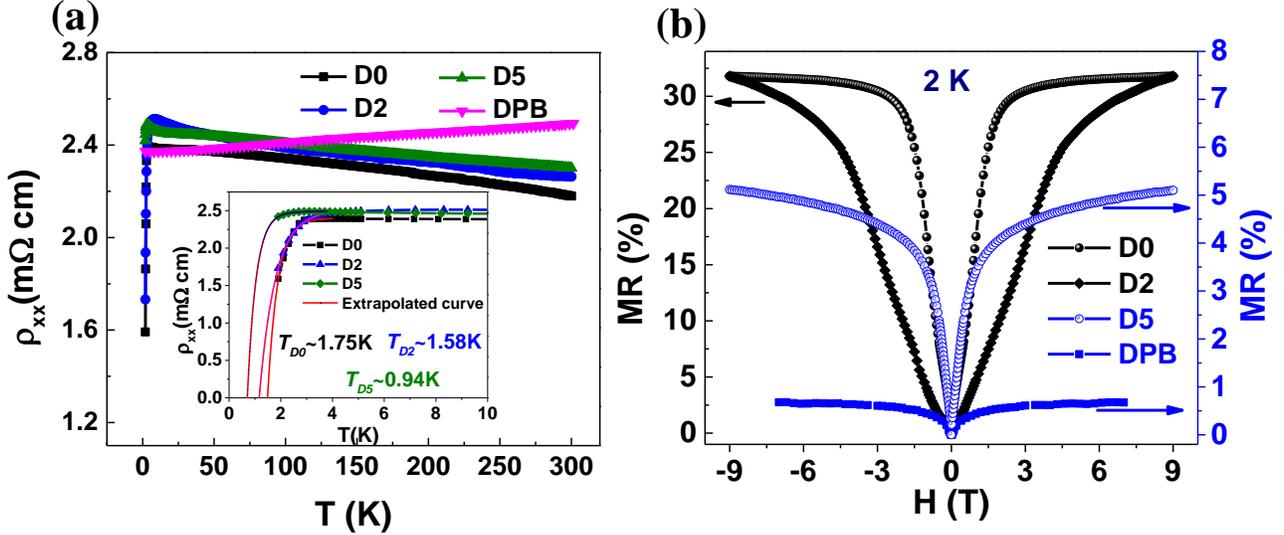

**FIG. 2.** (a) The $\rho_{xx}$ vs T plots for samples D0, D2, D5 and DPB in temperature range 1.9K ≤ T ≤ 300K; inset shows the extrapolated curves at temperature below 1.9K. (b) MR data of D0 (open black symbols), D2 (closed black symbols), D5 (open blue symbols) and DPB (closed blue symbols) at 2K.

The MR data of all samples share a common feature: a sharp rise in low-field resistance centered around zero-field at low temperatures which is the weak anti-localization (WAL) effect and is widely reported in topological insulators[32–35] and topologically non-trivial semimetals[8,36–40]. We have fitted the magnetic field dependence of magneto-conductance ($\Delta G_{xx}$) using well-known Hikami-Larkin-Nagaoka[41] model and extracted the fitting parameters $\alpha$ and $L_\Phi$, see eqn. S1 and Fig. S3 in supplementary section for detailed information. The fitted parameters $\alpha$ and $L_\Phi$ are plotted as a function of temperature from 1.9K ≤ T ≤ 10K in Fig. 3. Red horizontal lines and blue curves correspond to $\alpha = -0.5$ and temperature-dependent power law fitting to $L_\Phi$, respectively. We notice one interesting fact: the value of $\alpha$ at or below 2 K is much lower than -0.5. Noting that $\alpha$ is a measure of the number of coherent transport channels and a value of -0.5 describes a single surface conduction channel, a decrease in the value of $\alpha$ denotes more contribution from bulk conduction channel. Interestingly, the minimum value of $\alpha$ decreases with decrease in Dy content which is also correlated with the superconducting transition temperature; D0 with a superconducting transition temperature of 1.75 K showing a significantly lower $\alpha$ at 2 K compared to DPB sample, whose onset of superconductivity is well below 1 K. Possibly, this implies that the onset of superconductivity increases the bulk contribution to the conduction channels. The value of $\alpha$ approaches to -0.50±0.05 in temperature range 3K≤ T ≤8K indicating the presence of a single surface conduction channel in all the samples in this temperature range. The coherence length ($L_\Phi$) here is the characteristic length of samples which defines the average distance up to which electrons maintain their phase coherence. The $L_\Phi$ measured at 2 K increases with Dy doping - 156nm, 174nm, 226nm and 450nm for D0, D2, D5 and DPB respectively, (Fig. 3). The $L_\Phi$ decreases with increase in temperature for all samples and temperature dependent power law is fitted to $L_\Phi$ values as shown by the blue curves in Figs. 3(a-d). The value of the exponent is very close to -0.5 above the onset transition temperatures (≥ 2.5K), in all these samples showing they are in the 2D limit[32,35,42].

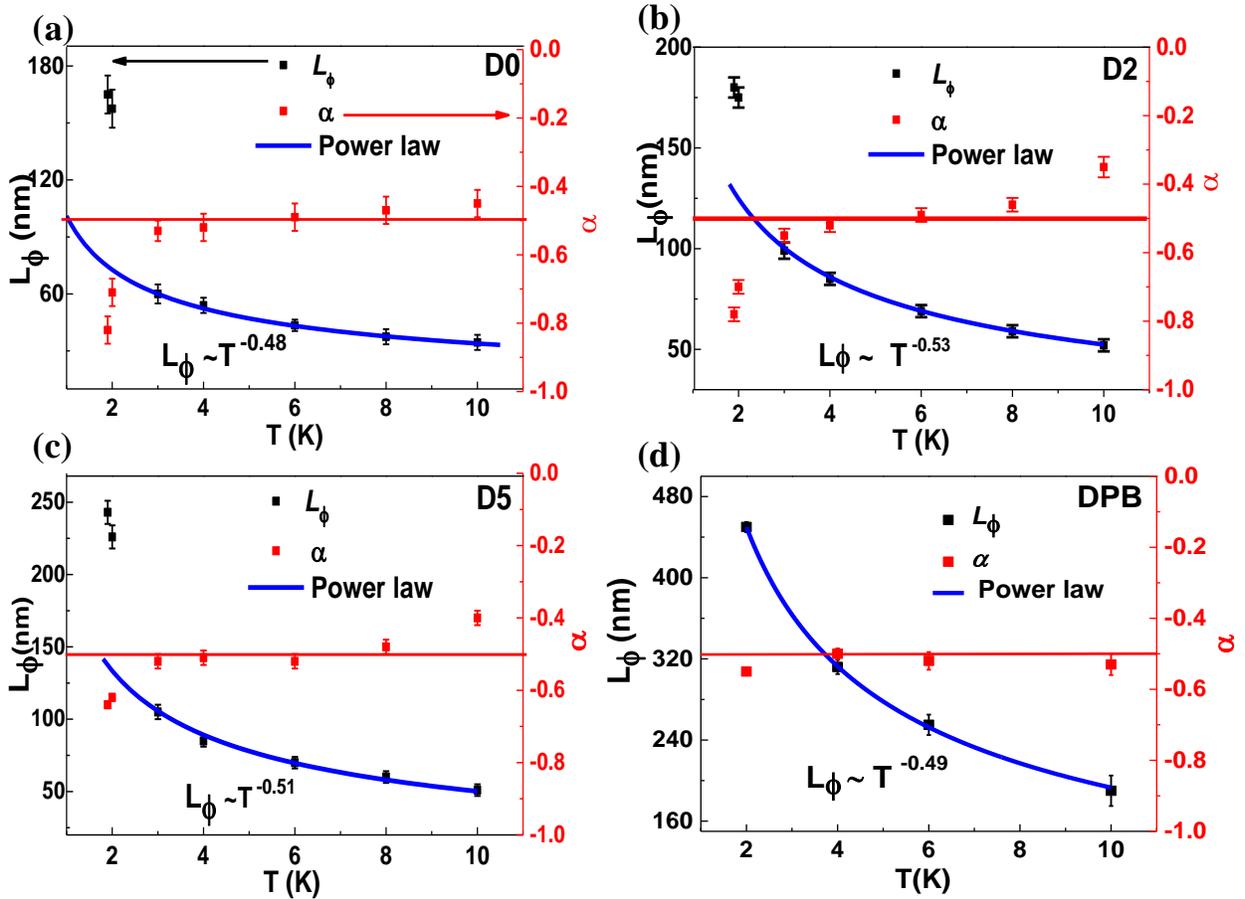

**FIG. 3.** Temperature dependence (1.9K≤ T ≤10K) of HLN fitting parameters $\alpha$ and $L_\Phi$ for samples (a) D0, (b) D2, (c) D5 and (d) DPB.

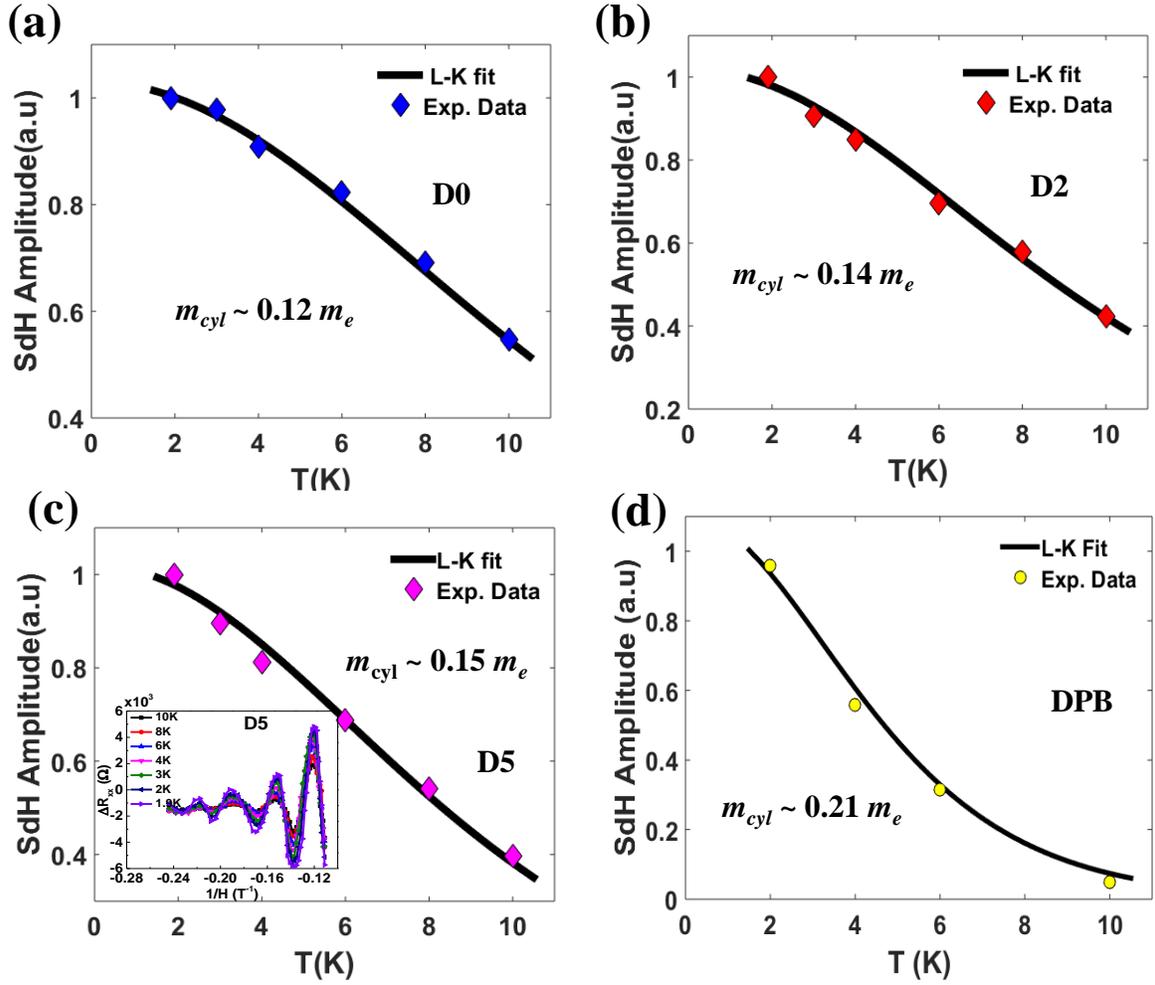

**FIG. 4.** Fitting of SdH oscillations amplitude to the thermal damping expression of Lifshitz-Kosevich equation for samples (a) D0, (b) D2, (c) D5 and (d) DPB. Inset of (c) shows the background subtracted SdH oscillations for sample D5.

High field MR data shows resistance oscillations known as the Shubnikov-de Haas (SdH) oscillations[43]. The background subtracted oscillatory part of $R_{xx}$ ($\Delta R_{xx}$) having periodic maxima and minima as a function of 1/H is shown in inset of Fig. 4(c) for D5 sample and Fig. S4. Clear SdH oscillations are observed up to 10K and amplitude of oscillations decreases with decreasing magnetic field. Single SdH frequency $f_{SdH}$ ~34T[9], 40T, 45T and 88T[25] are obtained from the FFT of the SdH oscillations for samples D0, D2, D5 and DPB respectively. The $k_F$ ~0.032Å$^{-1}$, 0.035Å$^{-1}$, 0.037Å$^{-1}$ and 0.05Å$^{-1}$ is estimated to increase with increasing LC % for D0, D2, D5 and DPB samples, respectively. The sheet carrier concentration ($n_s$)~8.21×10$^{11}$cm$^{-2}$, 9.69×10$^{11}$cm$^{-2}$, 10.89×10$^{11}$cm$^{-2}$ and 19.89×10$^{11}$cm$^{-2}$ is calculated for D0, D2, D5 and DPB respectively, which shows a clear increase with increasing Dy doping.

The cyclotron effective mass ($m_{cyl}$) of carriers is estimated by fitting amplitude of minima/maxima of oscillations to the thermal damping factor $R_T$ of eqn. S2 (see supplementary section). Figures 4(a-d) show the fittting and the extracted $m_{cyl}$ values of 0.12 $m_e$, 0.14 $m_e$, 0.15 $m_e$ and 0.21 $m_e$ ($m_e$ is the electron mass) for D0, D2, D5 and DPB respectively. Clearly, addition of Dy progressively increases the $m_{cyl}$ of the electrons which implies a decrease in the slope of the linear E-k band dispersion.

We estimate Dingle temperature ($T_D$)=5K, 4K, 3.5K and 5K for samples D0, D2, D5 and DPB respectively from the Dingle plots (see Figure S6). Various band structure parameters are extracted from analysis of SdH oscillations, see Table S2. From these values listed in the Table, it is clear that the very high values of the mobility ($\mu_S$) in these films together with low $m_{cyl}$ indicate dominant surface transport. The $\mu_S$ values for all the films are substantially larger in comparison to other rare based RPdBi HH alloys[9,24–26,36] and makes these samples suitable candidates for applications as functional materials for spintronic devices. Metallicity parameter ($k_Fl$) can give a practical indicator for the degree of metallic behavior of a system and it is found to be equal to 25, 30, 38 and 34 for D0, D2, D5 and DPB samples respectively. These high values indicate large surface state conductivities and are comparable to $Bi_2Te_2Se$[45].

The phase factor $\beta$ of SdH oscillations is calculated as $\beta$=0.46±0.1, $\beta$=0.41±0.1, $\beta$=0.43±0.1 and $\beta$=0.47±0.1 which correspond to Berry phase = 0.92π±0.1, 0.82π±0.1, 0.86π±0.1 and 0.94π±0.1 for samples D0, D2, D5 and DPB respectively using the well-known method of Landau level fan diagram[9,24,36,44], see Fig. S5. Ideally, Berry phase should be equal to π for exact linear E-k dispersion and it shifts if the bands deviate from linear dispersion by an offset and/or due to the Zeeman coupling of electron spin in presence of external magnetic field[46,47]. Observation of very small $m_{cyl}$ and high mobility with Berry phase ~ π indicate the presence of relativistic fermions in topologically non-trivial surface states.

Figure 5 summarizes our main findings with Dy doping which represents the correlation between *f* electron doping with crystal structure, electronic structure parameters and $T_C$. Figure 5 (a) shows that the LC % (black points) increases with Dy doping implying an increase in the lattice density. This is also confirmed from the XRR data as discussed earlier. This variation of lattice parameter usually leads to substantial modification of the band structure near the Fermi energy in HH systems[5,6]; here we observe that it leads to a substantial increase of the Fermi surface area (Fig. 5(a), green points) and an increase in the $m_{cyl}$ (Fig. 5(a), blue points). This demonstrates that doping at the rare-earth site offers a unique way to modify the electronic structure of parent YPdBi alloy and selectively vary the corresponding band structure parameters.

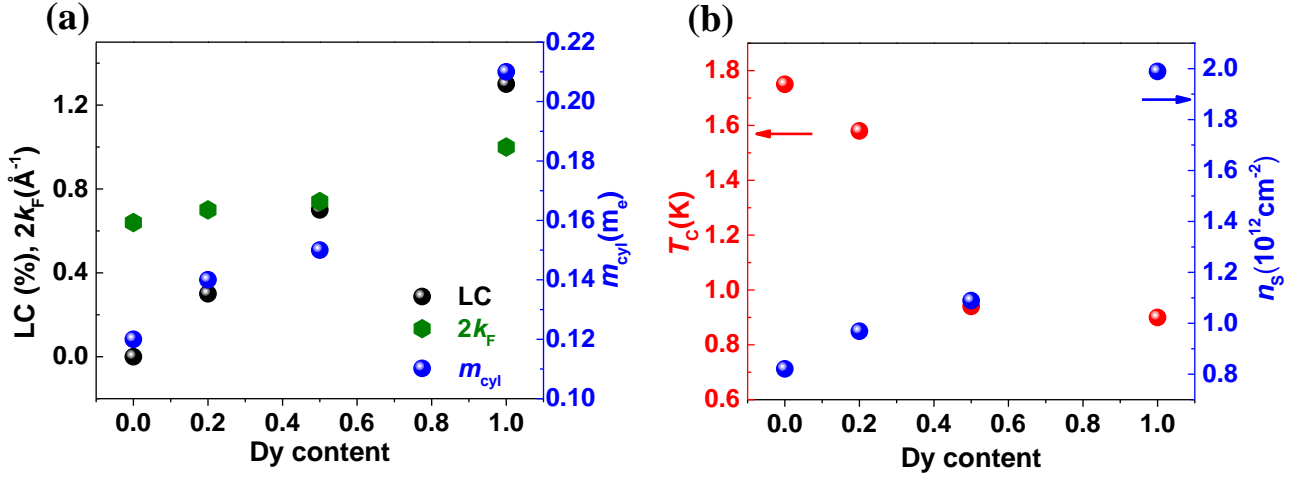

**FIG. 5**. Variation of electronic structure parameters (a) $k_F$, LC %, $m_{cyl}$ (b) $T_C$ and $n_S$ with Dy content. $T_C$ for pure DyPdBi sample is taken from ref. [2].

From Fig. 5(b) we observe that although the $n_s$ increases with increasing Dy doping, the $T_C$ is suppressed. This indicates an active role of the local moments arising from the 4f electrons of the $R^{3+}$ ion. In fact, an uniform linear suppression of $T_C$ as a function of the local moment strength (quantified by the de Gennes scaling factor) has been reported[2] for bulk RPdBi crystals. In our doped thin film system it indicates that with increasing doping of Dy, which has a non-zero de Gennes factor unlike Y, there is an increased coupling between the localized Dy spins and the conduction electrons via the Ruderman-Kittel-Kasuya-Yosida interaction. This naturally leads to an increased Cooper pair breaking effect arising from Dy local moments leading to a linear suppression of $T_C$ with Dy doping. A closer look at the $T_C$ variation (red points) shows that it somewhat deviates from this linear behaviour. One possible reason could be the simultaneous increase of the $n_s$ with Dy doping which possibly increases $T_C$ and in combination with the antagonistic pair breaking effect, leads to the deviation from linear suppression of $T_C$.

In summary, we conducted systematic studies on the structural, electrical and magneto-transport measurements of the *4f* electron doped $Y_{1-x}(Dy)_xPdBi$ (*x*=0, 0.2, 0.5, 1) strained thin films. The observations of 2D WAL, non-trivial Berry phase together with high mobility of low effective mass carriers indicate the surface dominated transport in these strained thin films. We observe a suppression of the onset of superconductivity with increasing Dy doping most likely arising from the enhanced local moment of the 4*f* electrons in Dy. The *f* electron doping provides a unique way to selectively tune the electronic structure of YPdBi alloy which are not only useful for designing future spintronics and quantum computing devices but also provides a pathway to systematically study the coexistence of competing superconducting and magnetic orders.

**Acknowledgements**

Authors would like to thank the NRF, CRF, PPMS and SQUID (Department of Physics) IIT Delhi for providing characterization facilities. NB and RC acknowledge the financial assistance received from SPARC proposal #754. NB acknowledges funding from the EPSRC through EP/S016430/1.

# Supplementary Information

**XRR fitting parameters:**

**Table S1.** Summary of XRR simulated parameters, here $t_{Ta}$ and $t_{YDPB}$ are the thickness of Ta and $Y_{1-x}(Dy)_xPdBi$ layers, $\sigma$ is the interface roughness, $\rho_{Ta}$ and $\rho_{YDPB}$ are the densities of Ta and $Y_{1-x}(Dy)_xPdBi$ layers, respectively.

| Sample | $t_{Ta}$ (nm) | $\rho_{Ta}$ (g/cc) | $\sigma$ (nm) | $t_{YDPB}$ (nm) | $\rho_{YDPB}$ (g/cc) |
|---|---|---|---|---|---|
| D0 | 5.1±0.09 | 14.78±0.3 | 1.2±0.02 | 31.1±0.06 | 8.45±0.2 |
| D2 | 4.9±0.05 | 15.11±0.1 | 0.8±0.01 | 29.2±0.05 | 8.89±0.1 |
| D5 | 5.2±0.08 | 15.19±0.2 | 1.1±0.02 | 28.2±0.08 | 9.31±0.2 |
| DPB | 5.1±0.08 | 14.86±0.2 | 1.5±0.02 | 32.7±0.09 | 9.71±0.2 |

**Field dependent resistivity:**

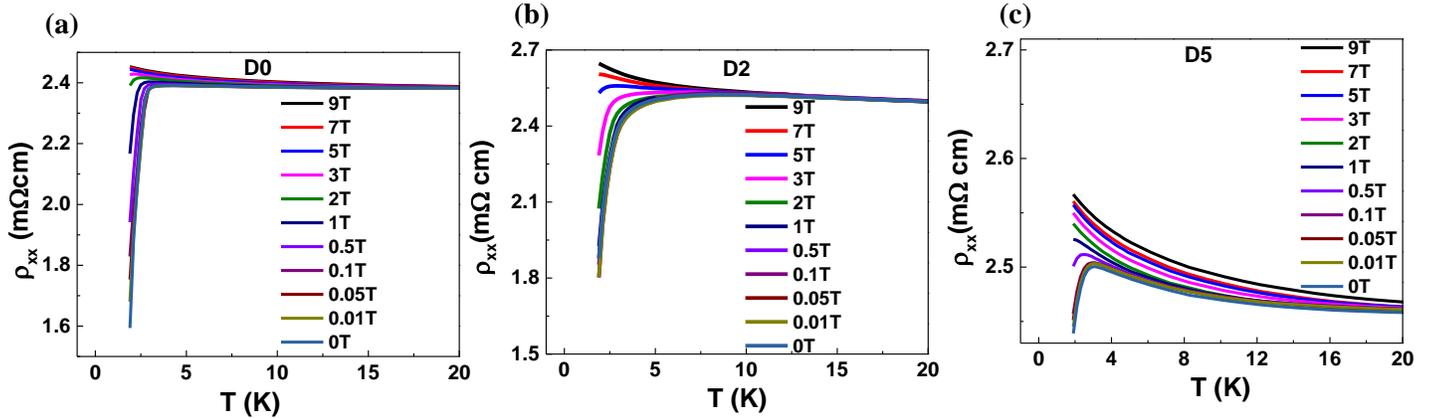

**FIG. S1.** Longitudinal resistivity measured in presence of external magnetic field (H=0 T to 9T) for samples (b) D0, (c) D2 and (d) D5.

**Magneto-resistance:**

The MR% at an applied field H denoted by MR(H), is calculated using the relation $[\rho_{xx}(H)-\rho_{xx}(0)]/\rho_{xx}(0) \times 100\%$, here $\rho_{xx}(H)$ and $\rho_{xx}(0)$ are the longitudinal resistivities recorded at field values of H and zero respectively

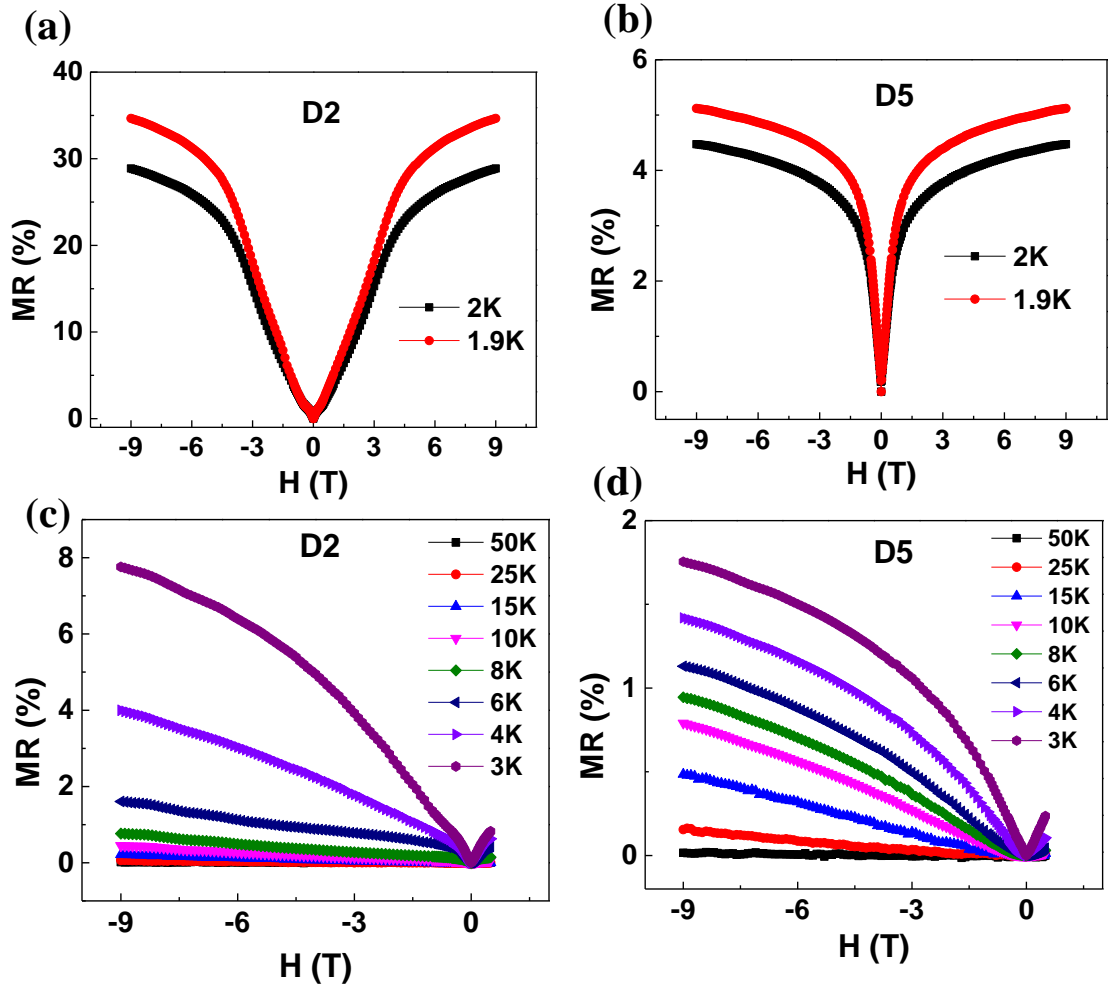

**FIG. S2.** (a-b) MR at 1.9K and 2K for samples D2 and D5. (c-d) MR in temperature range 3K ≤ T ≤ 50K for samples D2 and D5.

**HLN fitting to Weak-antilocalization :**

The reason of weak-antilocalization effect is as follows: the carrier paths in topological materials are protected by time reversal symmetry and these carriers have Berry phase = π associated with them. This Berry phase is the order parameter used to describe the non-trivial band structure of topological materials [1,2]. Due to π phase difference between these paths they interfere destructively after an adiabatic evolution around a non-magnetic impurity to preserve time reversal symmetry and a low resistance state is observed[3]. But as we apply external magnetic field this phase difference is shifted by an offset value and carrier paths start interfering constructively which leads to sharp increase in resistance.

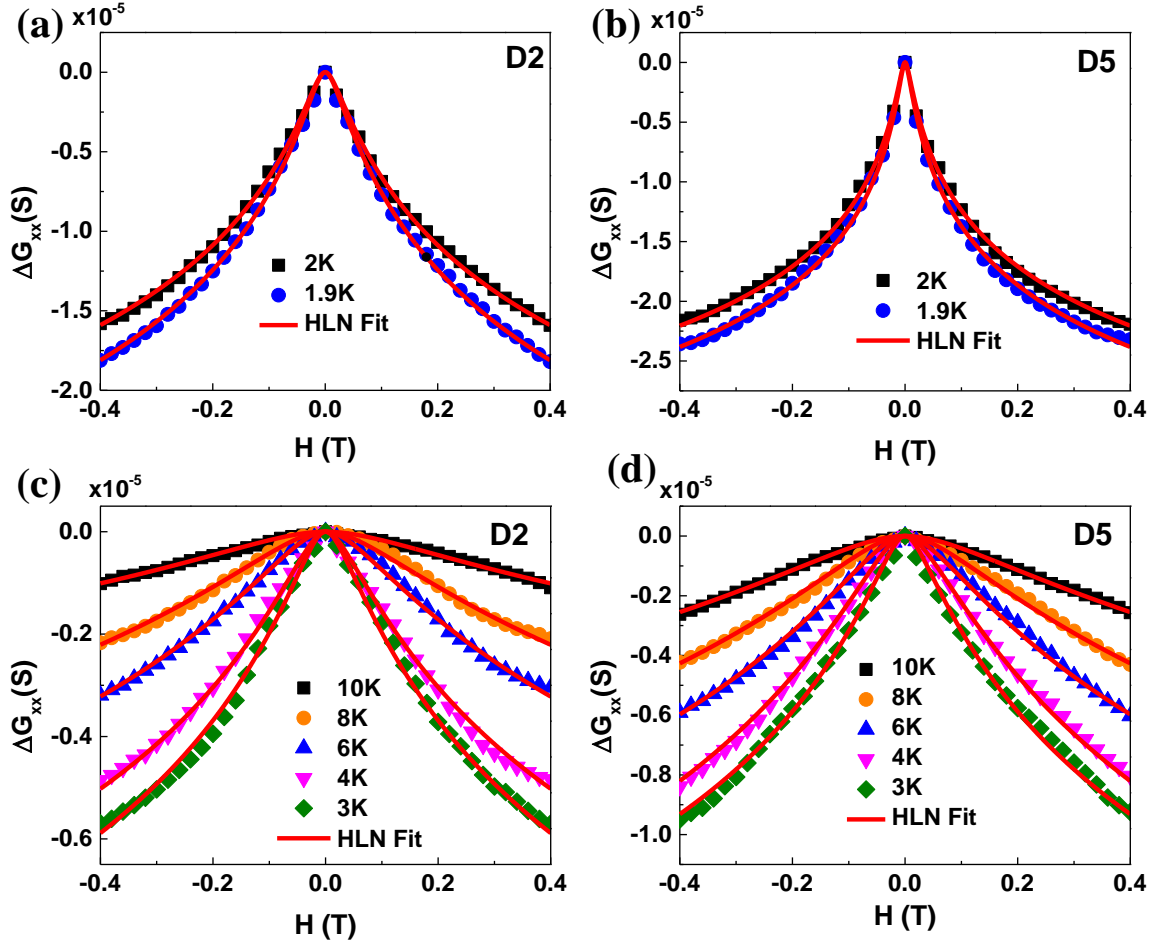

**FIG. S3.** HLN fitting (solid red line) of magneto-conductance data in -0.4T≤ H ≤+0.4T for samples D2 and D5 at (a-b) 1.9K and 2K; at (c-d) 3K≤ T ≤10K.

The WAL effect and its magnetic field dependence is described by the well-known Hikami-Larkin Nagaoka[4] model using the following equation:

$$\Delta G_{xx} = \gamma \left[ \ln \frac{H_\varphi}{H} - \Psi\left(\frac{1}{2} + \frac{H_\varphi}{H}\right) \right] \quad (S1)$$

here, $\gamma = \frac{\alpha e^2}{2\pi^2 \hbar}$ and $\Delta G_{xx} = G_{xx}(H) - G_{xx}(H=0)$ is magneto-conductance, $\Psi$ is the digamma function, $H_\varphi = \frac{\hbar}{4eL_\varphi^2}$, $\hbar$ and $e$ are the reduced Planck's constant and charge of electron, respectively. In this equation $\alpha$ and $L_\Phi$ are the fitting parameters, $L_\Phi$ is the phase coherence length of the carriers. The pre-factor $\alpha$ and temperature dependence of $L_\Phi$ provides information about the dimensionality of the topological system with $\alpha = -0.5$ and $L_\Phi \propto T^{-0.50}$ indicating 2D topological system while $\alpha = -1$ and $L_\Phi \propto T^{-0.75}$ indicate 3D topological system[5].

**Shubnikov-de Haas oscillations:**

The energy states of materials get modulated periodically in presence of high magnetic fields and oscillations are observed in resistivity which varies as periodic function of inverse magnetic field[5,6]. The information about Fermi surface and electronic structure of material can be extracted from the analysis of SdH oscillations. The SdH oscillations are well-described by Lifshitz-Kosevich (L-K) equation[7,8]:

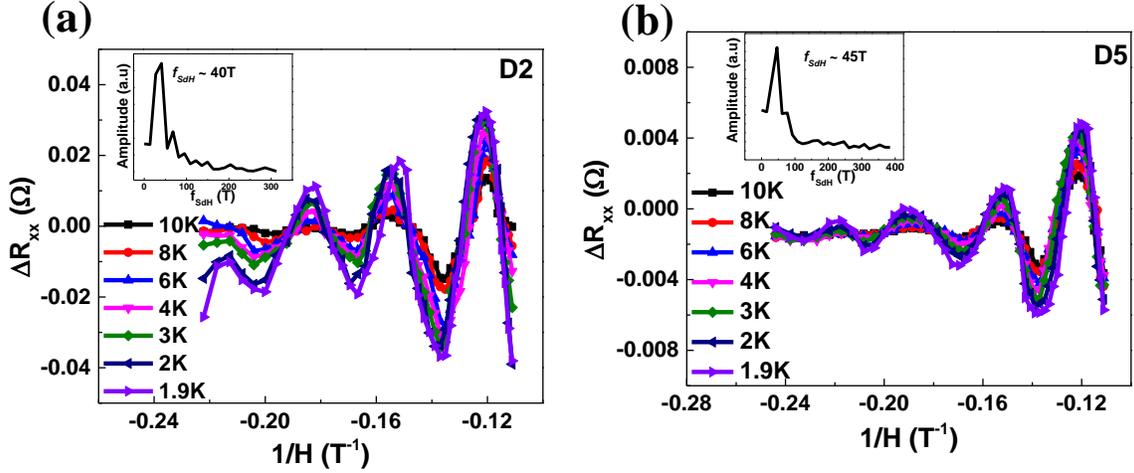

**FIG. S4.** SdH oscillations for T≤10K (a) D2 and (b) D5. Insets show FFT of SdH oscillations.

$$\Delta R_{xx} \propto R_T\, R_D R_S\, cos\left[2\pi\left(\frac{f_{SdH}}{H_n} - \frac{1}{2} + \beta\right)\right] \qquad (S2)$$

here $R_T$, $R_D$ and $R_S$ are the thermal, Dingle and spin damping factors, respectively. The dimensionality of the Fermi surface and other band structure parameters can be estimated from fitting of SdH data to above different factors of L-K equation. In the cosine function $f_{SdH}$ is the frequency of SdH oscillations which is related to cross sectional area of Fermi surface ($A_F$) using Onsager relation[8] as $f_{SdH} = (\frac{h}{4\pi^2 e})A_F$, $h$ is the Planck's constant. The Fermi surface cross sectional area ($A_F$) can be directly obtained from the SdH oscillation frequency ($f_{SdH}$) using the Onsager relation $f_{SdH} = (\frac{h}{4\pi^2 e})A_F$. Assuming a circular Fermi surface, the Fermi vector $k_F$ can be estimated from the relation $A_F = \pi k_F^2$. The $k_F$ is related to the sheet carrier concentration ($n_s$) via the relation $n_s = \frac{k_F^2}{4\pi}$ for 2D Fermi surface systems. The $H_n$ describe the magnetic field corresponding to the n$^{th}$ Landau level and the phase factor $\beta$ is related to Berry phase as = $2\pi\beta$. The topologically non-trivial semimetals have linear band dispersion around topologically protected nodes with β=1/2 (see Fig. S6) whereas β=0 for parabolic dispersion in case of trivial materials[5,6]. The broadening of Landau levels due to electron scatterings and sample inhomogeneity also damp the amplitude of oscillations and $R_D$ factor in equation S2 accounts for this damping phenomenon.

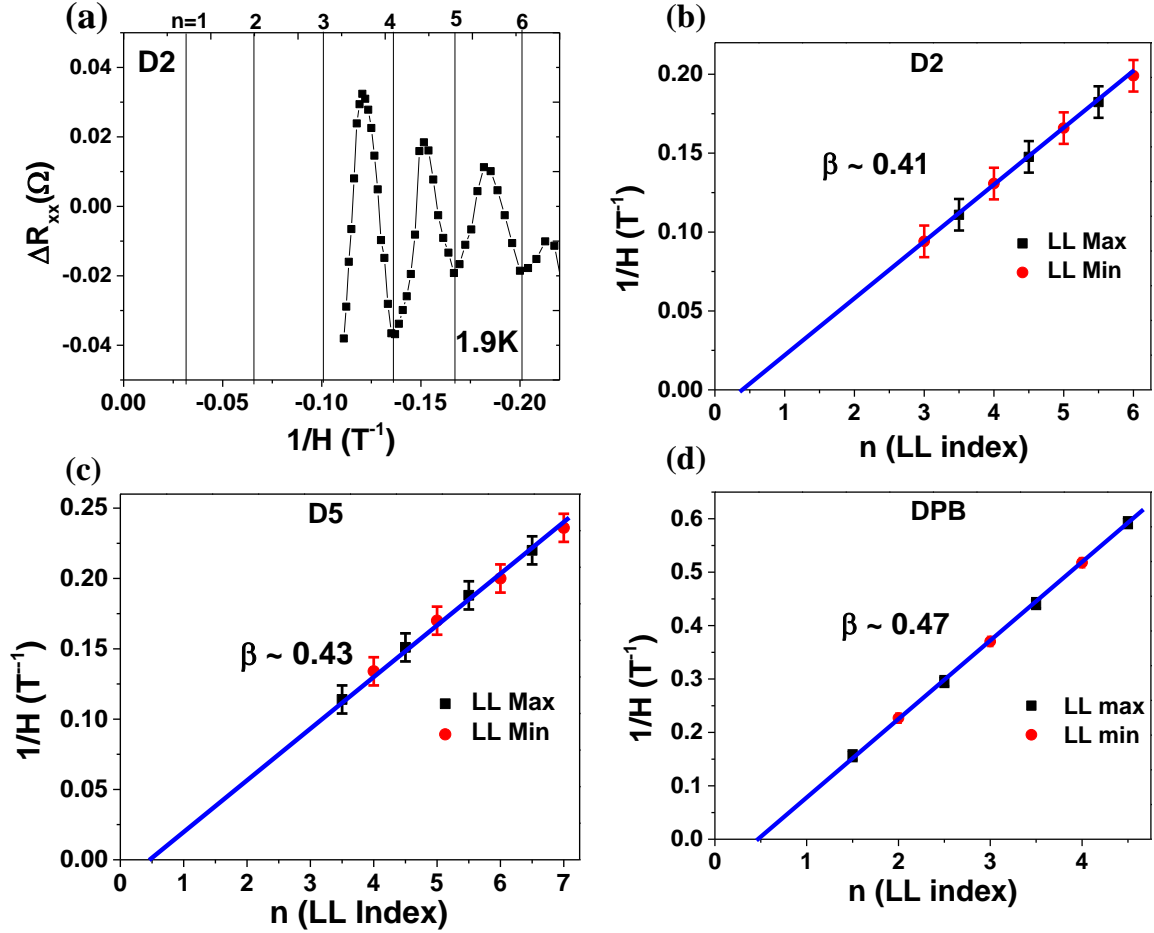

**FIG. S5.** Landau level fan diagram for sample (a-b) D2, (b) D5, and (d) DPB.

With increase in temperature the amplitude of SdH oscillations decreases and $R_T$ term in equation 2 accounts for this thermal damping. The cyclotron effective mass ($m_{cyl}$) of carriers can be estimated from SdH data by fitting amplitude of minima/maxima of oscillations to the thermal damping factor $R_T = \frac{2\pi^2 \kappa_B T/\Delta E_n(H)}{\sinh(2\pi^2 \kappa_B T/\Delta E_n(H))}$. In this equation $\Delta E_n(H)$ is the energy gap between two consecutive quantized Landau levels in presence of H and $\Delta E_n(H)$ is the fitting parameter in this equation. By using the relation $\Delta E_n(H) = \hbar eH/m_{cyl}$, we extracted the effective mass of carriers for all samples. Fermi velocity ($v_F$) and position of Fermi level ($E_F^S$) from the linear dispersion nodes are related to $k_F$ as $v_F = \hbar k_F / m_{cyl}$ and $E_F^S = (\hbar k_F)^2 / m_{cyl}$. The relaxation time, mobility and the mean free path of carriers are estimated from $T_D$ using relations $\tau = \hbar / 2\pi T_D \kappa_B$, $\mu_S = e\tau / m_{cyl}$ and $l = v_F \tau$, respectively see Table S2

**Dingle plots:**

The Dingle temperature ($T_D$) of samples can be calculated from SdH data by using the $R_D$ term where the slope of linear fit to $ln\{(\Delta R_{xx}(T)/\Delta R_{xx}(0)) H \sinh(2\pi^2 \kappa_B T / \Delta E_n)\}$ vs 1/H data points is used to estimate the $T_D$. This $T_D$ is the excess temperature required to fit the data and arises due to electron scatterings [8].

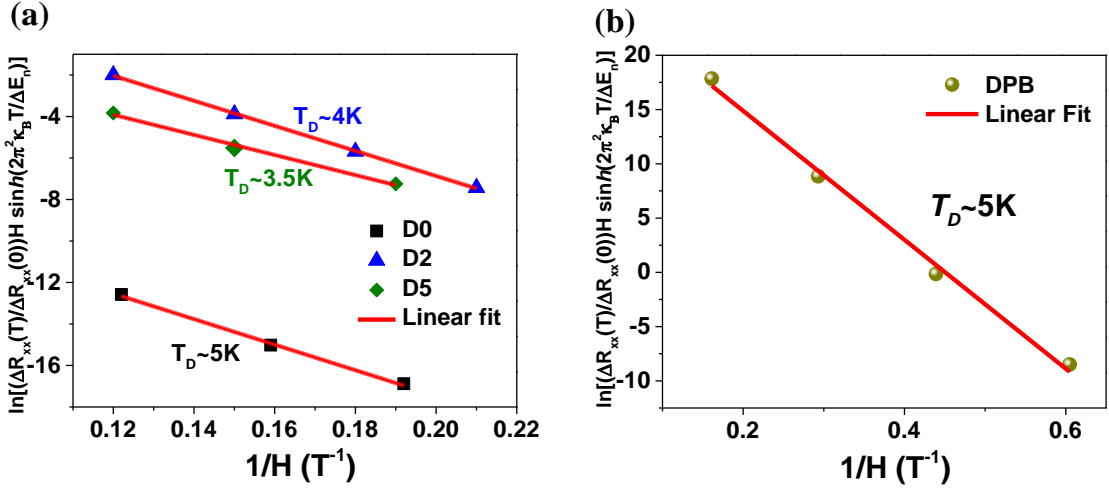

**FIG. S6.** SdH oscillations for T≤10K (a) D0, (b) D2,(c) D5 and (d) DPB. Insets show FFT of SdH oscillations.

**Table S2.** Parameters obtained from fitting of the SdH oscillations data to Lifshitz-Kosevich theory.

| | $f_{SdH}$ T | $T_D$ K | $n_s$ $10^{11}$ cm$^{-2}$ | $m_{cyl}$ $m_e$ | $k_F$ Å$^{-1}$ | $v_F$ $10^5$ ms$^{-1}$ | $E_F^S$ meV | $\tau$ $10^{-13}$ s | $l$ nm | $\mu_s$ cm$^2$ V$^{-1}$ s$^{-1}$ | $2\pi\beta$ $\pi \pm 0.1$ |
|---|---|---|---|---|---|---|---|---|---|---|---|
| **D0** | 34 | 5 | 8.21 | 0.12 | 0.032 | 3.21 | 68 | 2.43 | 78 | 3694 | 0.92 |
| **D2** | 40 | 4 | 9.69 | 0.14 | 0.035 | 2.88 | 66 | 3.04 | 87 | 3805 | 0.82 |
| **D5** | 45 | 3.5 | 10.89 | 0.15 | 0.037 | 2.84 | 69 | 3.47 | 98 | 4049 | 0.86 |

| | | | | | | | | | | |
|---|---|---|---|---|---|---|---|---|---|---|
| **DPB** | 88 | 5 | 19.89 | 0.21 | 0.05 | 2.78 | 91 | 2.43 | 67 | 2059 | 0.90 |